\documentclass[aps,prc,showpacs,twocolumn,preprintnumbers,floatfix, showkeys,tightenlines]{revtex4}
\usepackage{amsmath,amssymb,amsfonts}
\usepackage{graphicx}
\usepackage{color}
\allowdisplaybreaks

\newcommand{\be}{\begin{equation}}
\newcommand{\ee}{\end{equation}}
\newcommand{\bea}{\begin{eqnarray}}
\newcommand{\eea}{\end{eqnarray}}

\begin{document}



\title{Constraining the density dependence 
 of symmetry energy from nuclear masses}

\author{B. K. Agrawal$^{1}$, J. N. De$^{1}$, S. K. Samaddar$^{1}$,
G. Col\`o$^{2,3}$, and A. Sulaksono$^{4}$}

\affiliation{
             \textsuperscript{1} Saha Institute of Nuclear Physics,
Kolkata 700064, India\\
             \textsuperscript{2} Dipartimento di Fisica,
Universit\`a
degli Studi di Milano, via Celoria 16, I-20133 Milano, Italy\\
             \textsuperscript{3} INFN, sezione di Milano, via
Celoria
16, I-20133 Milano, Italy\\
             \textsuperscript{4}Departemen Fisika, FMIPA,
Universitas
Indonesia, Depok 16424, Indonesia}


\begin{abstract} 
Empirically determined values of the nuclear volume and surface symmetry
energy coefficients from nuclear masses are expressed in terms of
density distributions of nucleons in heavy nuclei in the local density
approximation. This is then used to extract the value of the symmetry
energy slope parameter $L$. The density distributions in both spherical
and well deformed nuclei calculated within microscopic framework with
different energy density functionals give $L = 59.0 \pm  13.0$ MeV.
Application of the method also helps in a precision determination of the
neutron skin thickness of nuclei that are difficult to measure accurately.

\end{abstract}

\pacs{21.65.Ef, 21.65.Mn, 21.10.Gv}

\keywords{ Symmetry energy, symmetry energy slope
parameter, nuclear matter, neutron skin }

\maketitle

 The nuclear symmetry energy measures the energy transfer in 
converting symmetric nuclear matter to the asymmetric one. The 
density dependence gives information on the isospin-dependent
part of the equation of state (EOS) of asymmetric nuclear matter.
The density content of symmetry energy is mostly encoded in 
the symmetry energy coefficient $C_v$, 
the symmetry slope parameter $L$ and the symmetry incompressibility
$K_{sym}$, all evaluated at the nuclear saturation density
$\rho_0$. Here $C_v(\rho )$ is the 
volume symmetry energy per nucleon of homogeneous
nuclear matter at density $\rho $,  
\begin{eqnarray}
C_v(\rho ) =[e(\rho ,X=1)-e(\rho,X=0)], 
\end{eqnarray}
\begin{eqnarray}
L~=3\rho_0 \left .\frac{\partial C_v(\rho )}
{\partial \rho }\right|_{\rho_0}, 
\end{eqnarray}
and 
\begin{eqnarray}
 K_{sym}~=9 \rho_0^2 
\left . \frac{\partial ^2C_v(\rho )}{\partial \rho ^2}\right|_{\rho_0}. 
\end{eqnarray}
In Eq.~(1), $e$ is the energy per nucleon and  $X =(\rho_n-\rho_p)/
(\rho_n+\rho_p)$ is the isospin asymmetry of the system.
The  parameters $C_v(\rho_0), L$ and $K_{sym}$ 
deem to be of fundamental importance
in both nuclear physics and astrophysics. The nuclear binding
energies, the position of the nuclear drip lines, the neutron skin
thickness or the neutron density distribution in neutron-rich
nuclei, $-$ all of these are known to have affectations from 
\cite{mye1,mye2,mol,bro,typ,fur,tod,dan} the symmetry
parameters so mentioned.
Many astrophysical phenomena also depend sensitively on the symmetry
slope parameter $L$. Most notable among them are the dynamical
evolution of the core collapse of a massive star and the associated explosive
nucleosynthesis \cite{ste1,jan}, the cooling of proto-neutron stars
through neutrino convection \cite{rob} or the radii of cold neutron
stars \cite{rob}. The nature and stability of phases within a neutron
star, its crustal composition, thickness and frequencies of crustal
vibrations \cite{ste2} also seem to be strongly influenced by the
symmetry energy and its density dependence. A glimmer of their
import could further be seen in relation to some issues of new
physics beyond the standard model \cite{sil,wen}.

Attempts on estimates of $L$ have been made in the last few years
from analyses of diverse experimental data, uncertainties still
linger, however. Pygmy dipole resonance \cite{car} in
$^{68}Ni$ and $^{132}Sn$ predicts a weighted average in the range 
$L$=64.8 $\pm $ 15.7 MeV, but giant dipole resonance in $^{208}Pb$
\cite{tri} points to a value of $L \sim $ 52 $\pm $7 MeV. Nucleon
emission ratios from heavy ion collisions \cite{fam} favor a  
value close to it, $L \sim $ 55 MeV, 
but isoscaling gives $L \sim $ 65 MeV \cite{she}
and isospin diffusion shifts the value further up, $L$ = 88 $\pm $ 25
MeV \cite{che,li}. Recently, analyzing nuclear energies within
the standard Skyrme-Hartree-Fock approach, Chen \cite{che1} 
brought  down $L$ to 52.5 $\pm $  20 MeV. However, from the
fit \cite{mol} of the experimental nuclear masses
to the calculated ones in the finite range droplet model,
the value of $L$ is found to be in the bound $L =$
70 $\pm $ 15 MeV. Inputs from astrophysical analyses, namely
neutron star masses and radii on the other hand constrain $L$
to 43 $ < L < $ 52 MeV \cite{ste3}. 

The neutron skin thickness $R_{skin}$ (=$R_n-R_p$; $R_n$ and $R_p$
are the neutron and proton root-mean squared (rms) radii) of $^{208}$Pb
calculated in the droplet model or in self-consistent quantal
calculations with many different interactions was found to have a strong 
linear correlation with the density dependence of symmetry energy 
around saturation \cite{cen,war,bro,typ,fur,tod}. 
Analyzing the correlation systematics of nuclear isospin with neutron
skin thickness for a series of nuclei in the ambit of droplet model,
the Barcelona Group \cite{cen,war} found  a range for as $L =$ 75 $\pm$
25 MeV. The neutron skin thicknesses  used by them are measured in antiprotonic atom
experiments \cite{trz,jas}, they carry the unavoidable strong interaction-
related uncertainties. This  had its imprint on their extracted  value of
$L$. The Lead Radius Experiment on Pb (PREX) promises a model-independent
probe of its neutron-skin thickness, but experiment-related uncertainties
could not put the value of $R_{skin}$ of $^{208}$Pb in more precision
than 0.302 $\pm $.175 fm \cite{abr,hor}. This large uncertainty mires $L$
in wide limits.

Given that the neutron skin thickness is a strong indicator of the
nuclear isovector properties, the nuclear dipole polarizability
$\alpha_D$ has been suggested to provide a unique constraint on
$R_{skin}$ \cite{rei,pie}. 
The recent high resolution ($p,p^\prime $)
measurement \cite{tam} of $\alpha_D$ yields the neutron skin
thickness of $^{208}$Pb to be 0.168 $\pm $ 0.022 fm \cite{pie1}.
Quadrupole 
resonance energies may similarly be exploited to give estimates
of $R_{skin}$. 
Microscopic calculations \cite{roc} of these observables 
based on families of non relativistic
and covariant energy density functionals (EDF) point to a value
of $R_{skin}$ in close vicinity, $R_{skin}=$ 0.14 $\pm $0.03 fm for
$^{208}$Pb. Consequently,  value of the symmetry slope parameter
extracted from the GDR and GQR measurements turns out to be 
$L=$ 49 $\pm $ 11 and  37 $\pm $ 18 MeV, respectively.   

Nuclear masses are one of the most precisely experimentally determined 
quantities in nuclear physics. They provide information, through the 
liquid-drop mass systematics on the values of the volume and surface 
symmetry energy coefficients $C_v(\rho_0 )$ and $C_s$. 
The symmetry energy coefficient, $a_{\rm sym}(A)$, for a nucleus with mass
number $A$ can be expressed in terms of $C_v(\rho_0)$ and $C_s$ as,
\begin{eqnarray}
 a_{sym}(A)=C_v(\rho_0)-C_s A^{-1/3}.
\end{eqnarray}
An alternative form for $a_{sym}(A)$  has also 
been suggested \cite{dan1}, namely,
\begin{eqnarray}
 a_{sym}(A)=\frac{C_v(\rho_0)}{(1+\kappa A^{-1/3})},
\end{eqnarray}
where $\kappa =C_s/C_v(\rho_0)$. From analyses of $\sim $ 2000 nuclear
masses, Liu $\it et~ al. $ \cite{liu} obtained $C_v(\rho_0) =$31.1 $\pm $
1.7 MeV and $\kappa =$ 2.31 $\pm $ 0.38. Equating $a_{sym}(A)$ with 
$C_v(\rho_A)$, where $\rho_A$ is an equivalent density specific to 
mass $A$ and taking an ansatz for the density dependence of $C_v(\rho )$
as
\begin{eqnarray}
C_v(\rho )= C_v(\rho_0)(\frac {\rho }{\rho_0 })^\gamma ,
\end{eqnarray}
they obtain, choosing $\rho_A$ $\sim $ 0.1 fm$^{-3}$ for $^{208}$Pb,
$\gamma =$ 0.7 $\pm $ 0.1 and thence $L=$ 66 $\pm $13 MeV.
The density dependence of $C_v$ as given by Eq.~(6) is quite
consistent with the density dependence obtained from the nuclear
EOS with different interactions \cite{che,li,sam,fam,she}; 
the value of $\gamma $ varies though from interaction
to interaction. The 
choice for the value of the equivalent density has no firm 
theoretical basis, however.

A meticulous study \cite{jia} of the double differences of 'experimental'
symmetry energies were done very recently.  The double differences
in symmetry energies of neighboring nuclei has the advantage that
effects from pairing and shell corrections are well canceled out,
resulting in a compact correlation between the double differences
and the mass number of nuclei.  This yields values of $C_v(\rho_0 )$
and $C_s$ as 32.1$\pm $0.31 MeV and 58.91 $\pm $1.08 MeV, respectively,
values whose uncertainties are much smaller than those found previously.
Exploiting the empirical values of $C_v(\rho_0)$ and $C_s$, an exploration
was done earlier \cite{agr} with the ansatz given by Eq.~(6) to calculate
the equivalent density $\rho_A$ for the 'benchmark' spherical nucleus
$^{208}$Pb in the local density approximation \cite{sam}. Hereafter, we
refer to the choice of Eq.~(6) for the form of density dependence of
$C_v (\rho)$ as case I and the empirical values  $C_v(\rho_0)$ and $C_s$
as $C_v^0$ and $C_s^0$ respectively.

The saturation density was chosen as $\rho_0 =$ 0.155 $\pm $0.008
fm$^{-3}$; this covers the saturation densities that come out from
the EOSs of different Skyrme and relativistic mean-field (RMF) models.
The empirical information of the proton rms radius in $^{208}$Pb and
the correlation of the symmetry slope parameter with the neutron skin
thickness  of $^{208}$Pb, all deduced from a family of RMF energy density
functionals served to give a value of $\gamma \sim $ 0.664 $\pm $ 0.051,
and hence $L =$ 64 $\pm $ 5 MeV \cite{agr}. The neutron skin thickness
for $^{208}$Pb is found to be 0.188 $\pm $0.014 fm, the equivalent
density $\rho_A$, comes out to be 0.089 fm$^{-3}$ from the model.

To confirm the robustness of the procedure adopted in our recent
work \cite{agr}, it is imperative to see whether choice of forms of
density dependence of symmetry energy other than that given by Eq.~(6)
produce nearly the same results. It is also important to check if other
heavy nuclei, spherical or otherwise, collate the conclusions arrived with
this procedure.  In the present communication we use different functional
forms for the $C_v(\rho)$, employed very recently in Ref. \cite{tsa,dong}
in checking  the validation of  the relationship  between $C_v(\rho_0)$,
$L$ and $K_{\rm sym}$ using different effective interactions. Our
calculations are performed for both the spherical and well deformed
$^{208}$Pb and $^{238}$U nuclei, respectively.  The density distribution
of  nucleons corresponding to different neutron-skin thickness, required
for our calculations, are obtained using different energy density
functionals based on the RMF model. The choices for these energy density
functionals are exactly the same as in Ref. \cite{agr}.  

The form of $C_v(\rho )$ given by Eq.~(6) (case I) has some limitations
\cite{dong}.  With this functional form, $C_v(\rho_0 ), L$ and $K_{sym}$,
when calculated from Eqs.~(1) -(3) can be correlated as,
\begin{eqnarray}
C_v(\rho_0)=\frac{L}{3+K_{sym}/L}.
\end{eqnarray}
The results for the relative error $\Delta C_v/C_v$, obtained using
different RMF models, are displayed in Fig. 1. The $\Delta C_v$ for
the case I (black squares) is the difference in the lhs and rhs  of
Eq. (7). The values of $C_v$, $L$ and $K_{\rm sym}$ are calculated at
$\rho_0$ using Eqs. (1), (2) and (3), respectively.  The departure of
$\Delta C_v/C_v$ from zero is indicative of the inaccuracy involved
in expressing the functional dependence of symmetry energy as given
by Eq.~(6).

 \begin{figure}
\resizebox{3.0in}{!}{ \includegraphics[]{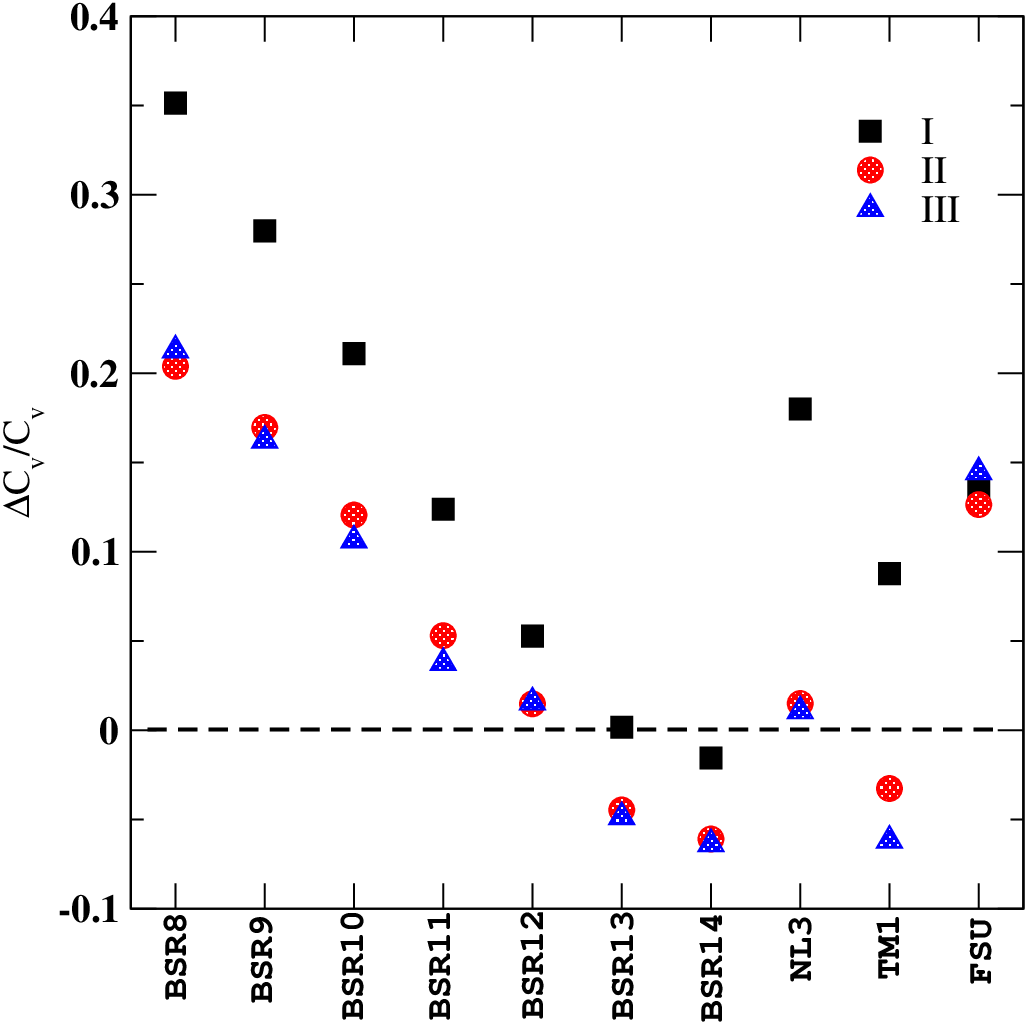}}
\caption{\label{fig:snm1} (Color online) The relative error $\Delta
C_v/C_v$ calculated at $\rho_0$ are shown for the various interactions
pertaining to the cases I, II, and III. The meaning of the three cases
are explained in the text.}
  \end{figure}
Another functional form for $C_v(\rho )$ seems more pertinent  \cite{tsa}:
\begin{eqnarray}
C_v(\rho )= C_k (\frac{\rho }{\rho_0})^{2/3} +(C_v(\rho_0 )-C_k) 
(\frac{\rho }{\rho_0})^\gamma.
\end{eqnarray}
The first term  in the rhs of the  above equation takes care
of the kinetic energy contribution to the $C_v(\rho)$ , the last term
comes from interactions. The value of $C_k$ is $C_k= (2^{2/3}-1)\times
\frac{5}{3}\frac {P_{F,0}^2}{2m^*} $ where $P_{F,0}$ is the Fermi
momentum corresponding to the saturation density $\rho_0$ and $m^*$
is the nucleon effective mass.

Inspired by the density-dependent M3Y (DDM3Y) interaction, 
the density dependence of $C_v(\rho )$ has also been shaped as 
\cite{muk},
\begin{eqnarray}
C_v(\rho )= C_k (\frac{\rho }{\rho_0})^{2/3} +C_1 
(\frac{\rho }{\rho_0})+C_2 (\frac{\rho }{\rho_0})^{5/3},
\end{eqnarray}
where $C_2=C_v(\rho_0)-C_1-C_k$.  The density dependence of $C_v(\rho )$
in the DDM3Y interaction, however, shows a different behavior as compared
to the RMF models.   Never the less, we use this functional form
also to check the robustness of our calculations.  The forms for
$C_v(\rho)$ depicted by Eqs.~(8) and (9) are hereafter referred to as case
II and case III. The three symmetry parameters are correlated as,
 \begin{eqnarray}
C_v(\rho_0 )= C_k +\frac{(L-2C_k)^2}{3L+K_{sym}-4C_k},
\end{eqnarray}
and
\begin{eqnarray}
C_v(\rho_0 )= \frac {L}{3}-\frac {K_{sym}}{15}+\frac {C_k}{5},
\end{eqnarray}
for the cases II and III respectively. As seen in Fig.~1, the values of
$\Delta C_v(\rho_0)$ for cases II (red circles) and III (blue triangles)
are quite similar for the interactions we have chosen; they lie closer
to zero in comparison to those in case I, except for the BSR13  and
BSR14 interactions.

From the above results, one can infer about the adequacy of the three
functional forms for modeling the density dependence of the symmetry
energy only at the saturation density.  In Fig. 2, we plot the values of
$C_v(\rho)$ over a wide density range 0 $\leq \rho \leq 3\rho_0$ which are
obtained using Eq.~ (1) for a few representative interactions.
These values of $C_v(\rho)$ are also compared with the ones obtained
by fitting them to the cases I, II and III.  In this density range,
one sees that the symmetry energy has a stiff density dependence for
NL3 \cite{lal} interaction, BSR10 \cite{agr1} and FSU \cite{tod}
display a softer dependence.  All the three functional forms (Eqs.
(6), (8), and (9)) are seen to give a moderatly good  representation of the
density dependence of symmetry energy for the specific RMF interactions
except at supranormal densities.  For the DDM3Y interaction, it is seen
that only case III (Eq.~(9))compares extremely well with $C_v(\rho )$.
This is so because of construction.

 \begin{figure}
\resizebox{3.0in}{!}{ \includegraphics[]{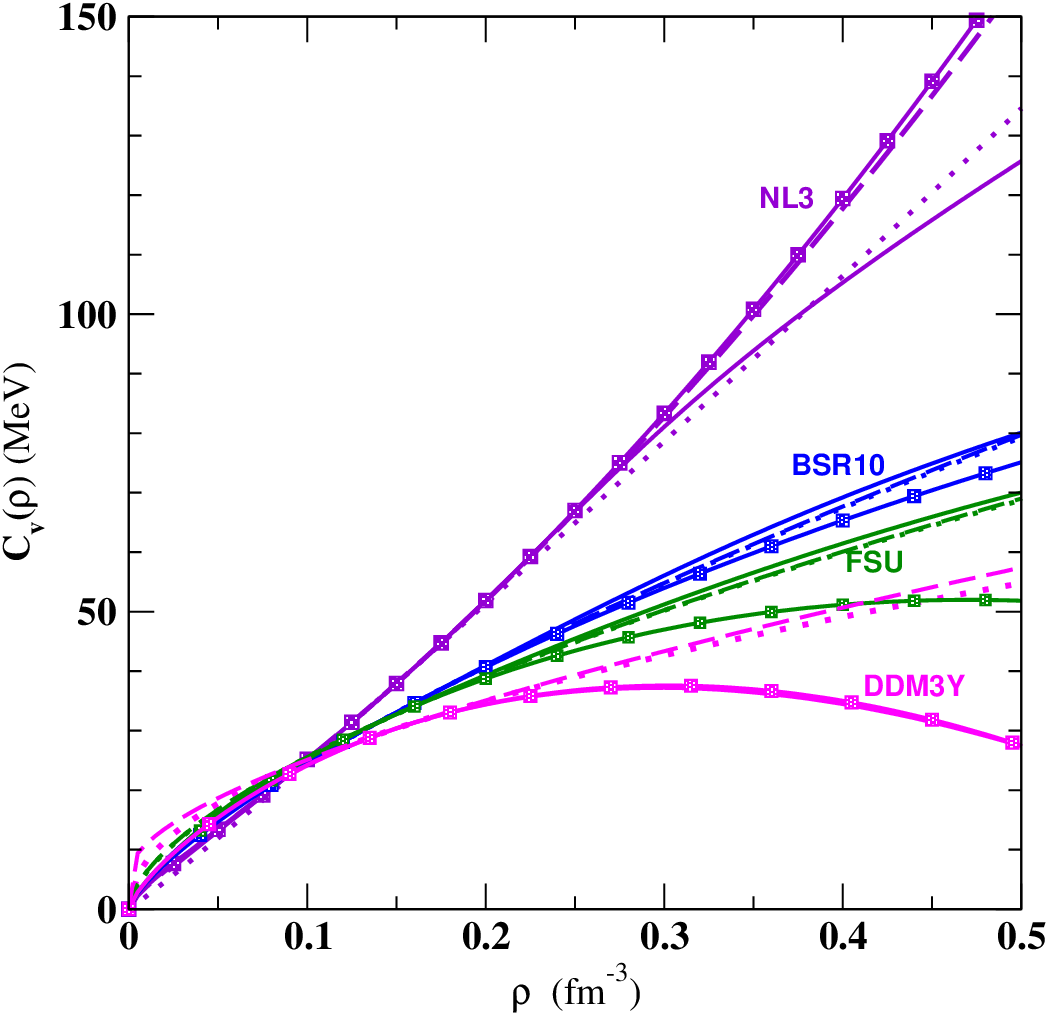}}
\caption{\label{fig:snm2} (Color online) Symmetry energy plotted as a
function of density for a few representative interactions.  The lines in
violet, blue, green and magenta colors correspond to NL3, BSR-10, FSU and
DDM3Y interactions, respectively.  The solid lines refer to $C_v(\rho
)$ calculated from Eq.~(1), the dotted, dashed and lines with square
symbols refer to case I, II and III, respectively.  For the BSR10 and
FSU, results for cases I and II overlap on each other. For the DDM3Y,
result for case III is almost indistinguishable from  the one calculated
using Eq.~(1).
  } \end{figure}

 To a good approximation, $C_v(\rho )$ is given as \cite{cen}
\begin{eqnarray}
C_v(\rho )=C_v(\rho_0)-L\epsilon +\frac{K_{sym}}{2}\epsilon ^2
\end{eqnarray}
where $\epsilon =(\rho_0 -\rho )/3\rho_0$. Since $a_{sym}(A)=C_v(\rho_A)$,
one immediately gets, from Eqs.~ (4) and (12), 
\begin{eqnarray}
C_s=A^{1/3}[L\epsilon_A-\frac{K_{sym}}{2}\epsilon_A^2],
\end{eqnarray}
where $\epsilon_A =(\rho_0 -\rho_A )/3\rho_0$. 
The symmetry coefficient $a_{sym}(A)(\equiv C_v(\rho_A))$, in the 
local density approximation can be calculated as,
\begin{eqnarray}
C_v(\rho_A)=\frac{1}{AX_0^2}\int d{\bf r} \rho {\bf (r)} C_v(\rho{\bf
(r)})[X{\bf (r)}]^2,
\end{eqnarray}
where $X_0 (= \frac{N-Z}{A} )$ is the isospin asymmetry of the nucleus,
$\rho {\bf (r)}$ is the sum of its neutron and proton densities and
$X{\bf (r)}$
is the local isospin asymmetry. Once the neutron-proton density
profiles in the nucleus are known, for case II say, with $C_v(\rho )$
as given by Eq.~(8), a chosen value of $\gamma $ gives $\rho_A$ and
hence $\epsilon_A$. The symmetry slope parameter $L$ and  $K_{sym}$
are  obtained from the density derivative of $C_v(\rho )$ as $L=2C_k+
3(C_v^0-C_k) \gamma $, the symmetry incompressibility $K_{sym}$
is determined from Eq.~(10) using this value of $L$ and the input
value of $C_v(\rho_0)$ (=$C_v^0$).  Only when Eq.~(13) is satisfied
(with $C_s=C_s^0$), one obtains the desired solution $\gamma $, thence
$L$, $K_{sym}$ and $\rho_A$.  These can, likewise be determined for
Case I. In a similar fashion, for case III, $C_1$ of Eq.~(9) can be
determined iteratively to give $L= 5C_v^0-3C_k-2C_1 $ and $K_{sym}=
10C_v^0-12C_k-10C_1$.
 \begin{figure}
\resizebox{3.0in}{!}{ \includegraphics[]{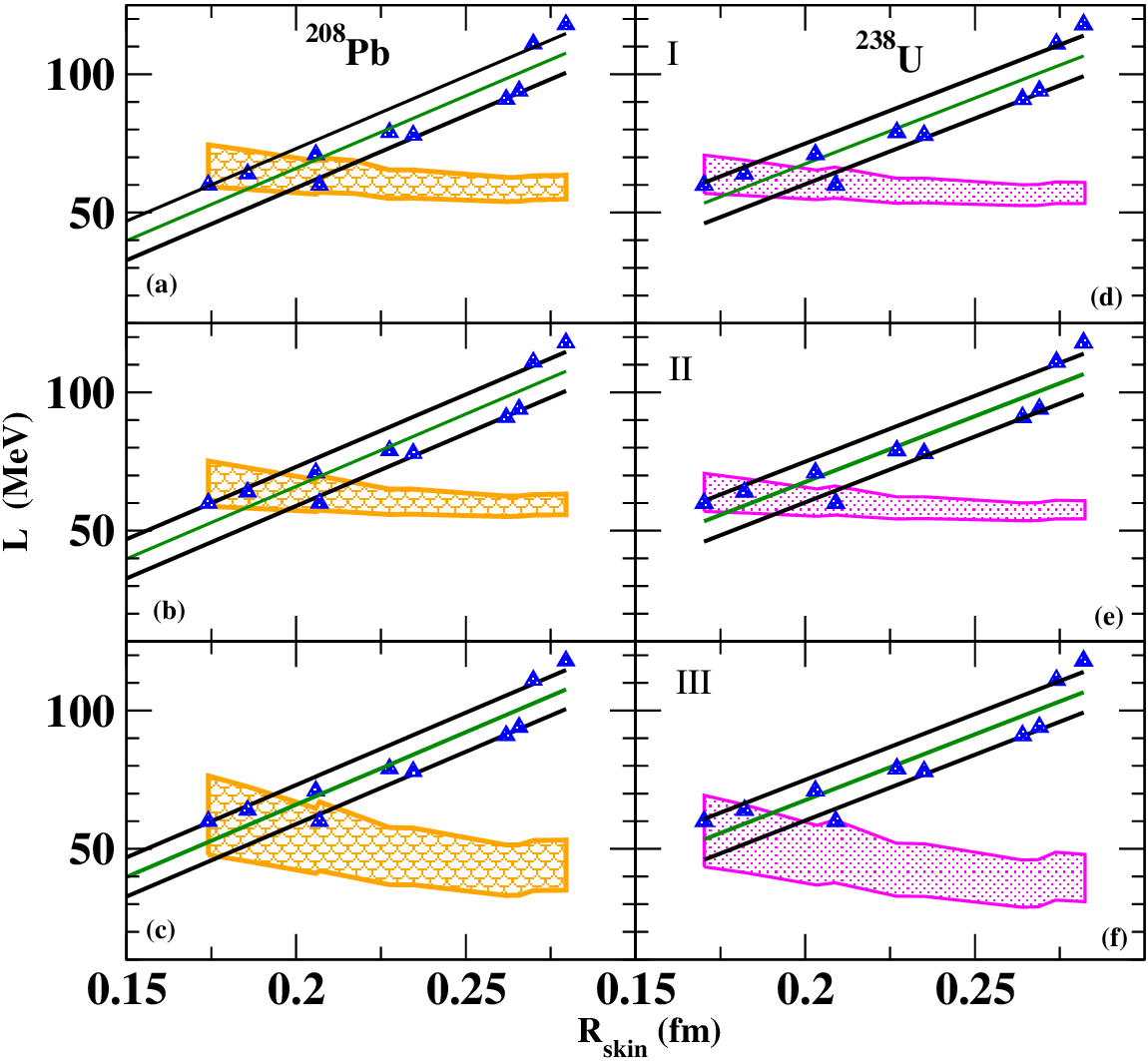}}
\caption{\label{fig:snm3} (Color online)  Symmetry slope parameter $L$
calculated using Eq.~(2) plotted as a function of $R_{skin}$ of $^{208}$Pb
and $^{238}$U evaluated with the different RMF interactions are shown
by the blue triangles. The green lines with envelopes of slanted black
lines  refer to the least-squared fits
to them with the spread-out error. The shaded regions in the left and
right panels represent the envelope of possible $L$ values calculated in
case I, II and III. The intersection of the slanted envelopes  and the shaded
regions depict the acceptable window for the values of $L$ and $R_{skin}$
for the cases concerned.  }
  \end{figure}

The procedure as described works best for heavy nuclei where volume
effects predominate over surface effects.  In Ref. \cite{agr}, for case
I, $^{208}$Pb was chosen as a representative nucleus. It usually serves
as a benchmark for comparison with extracted bulk nuclear properties.
The present calculations along with the spherical nucleus $^{208}$Pb
are repeated for another heavy  and deformed nucleus $^{238}$U to see
if the extracted informations from the two nuclei are in consonance.

We have used the interactions BSR8-BSR14 \cite{agr1}, FSU
\cite{tod}, NL3 \cite{lal}, and TM1 \cite{sug} to generate the neutron
and proton density profiles of the two nuclei in the RMF model. These
interactions in general  reproduce many experimental bulk properties of
finite nuclei and nuclear matter. The binding energies of $^{208}$Pb and
$^{238}$U are obtained in close experimental agreement. The calculated
values of the quadrupole deformation $\beta_2 = 0.26 - 0.28$ for
$^{238}$U from   the various RMF models considered here agree very well
with the experimental value $ \sim 0.29$.  The proton rms radii are also
reproduced nearly exactly. There is a wide variation in the neutron rms
radii, however; the calculated neutron skin thickness varies from $\sim
$ 0.17 fm to $\sim $ 0.28 fm.  In Fig.~3(a)-3(f), the symmetry energy
slope parameters evaluated with these interactions using Eq.~(2) are
displayed as a function of the corresponding calculated $R_{skin}$ (blue
filled triangles) for $^{208}$Pb and $^{238}$U.  For both the nuclei,
an almost linear correlation of $L$ with $R_{skin}$ is observed. We
also use Eqs.~(6), (8), (9), (13), and (14) to calculate $L$ from these
functional forms (case I-III) by employing the microscopic nuclear
densities for $^{208}$Pb and $^{238}$U obtained within the RMF models
using the empirical constraints on the parameter sets $C_v^0$ (=32.1$\pm
$0.31 MeV), $C_s^0$ (=58.91 $\pm $1.08 MeV) and $\rho_0$ (=0.155 $\pm
$0.008 fm$^{-3}$). The so-calculated $L$ values are depicted in the
figures by a shaded region, the spread coming from the uncertainties
in the values of $C_v^0, C_s^0$ and $\rho_0$. These values display an
altogether different type of correlation of $L$ with $R_{skin}$, the
weak dependence coming from the imposed empirical constraints.

The change in $L$ with $R_{skin}$ as displayed by the filled triangles,
after least-squares fitting, gives an almost  linear correlation shown by
the green  straight lines passing through the triangles.  The correlation
coefficient for both $^{208}$Pb and $^{238}$U nuclei is $0.93$. The small
deviations from complete linearity yields the rms error on $L$ values
$\sim 7.2$ MeV.  The spread in the $L$ values for a given neutron-skin
thickness within the RMF models is contained within the  slanted black lines in
the Fig. 3.  Their intersection with the shaded region projects out
those value of neutron skin thickness  for the nuclei concerned and
also the density slope parameter $L$ that are commensurate with the
empirical windows for $C_v^0, C_s^0$ and $\rho_0$.  The values of the
symmetry slope parameter $L$ and the neutron skin thickness  $R_{skin}$
for the two nuclei and the three cases are summarized in Table I.
It is seen that for a particular functional form for $C_v (\rho )$
(case I-III) chosen, the value of $L$ is nearly independent of the
heavy system.  If results for all the three cases and the two nuclear
systems are put together, the value of $L$ spans the range  59.0 $\pm $
13.0 MeV. Case III, designed specifically for M3Y interaction shows
a distinctly different behavior for $C_v(\rho )$ at somewhat higher
density beyond saturation as shown in Fig.~2.  Remembering that in case
II (as opposed to case I), the density dependence of the symmetry energy
originating both from the kinetic and interaction parts are tied in,
it appears that case II is the most suitable choice for the functional
form of $C_v(\rho )$. There is not much difference in the values of $L$
deduced from the two cases though, this is because the exponent $\gamma $
in case I is not much different from the density exponent in the symmetry
kinetic energy.  There is some confusion as to whether the effective
nucleon mass $m^*$ or the bare mass $m$ should be used in evaluating
$C_k$, we find that effects of interaction implicitly hidden in $m^*$
are compensated by the corresponding change in $\gamma $ so that the
value of $L$ is nearly unchanged.

\begin{table}
\begin{center}
\caption{
The values of symmetry energy slope parameter $L$ and the neutron-skin
thickness 
$R_{\rm skin}$ obtained for different forms for the density dependence
of $C_v(\rho)$, labelled as case I, II and III.}
\begin{tabular}{ccc|cc}
\hline\hline
 & & $^{208}Pb$&  & $^{238}U$\\
\hline
  & L (MeV)& $R_{\rm skin}$(fm)  & L (MeV)& $R_{\rm skin}$(fm)\\
\hline
Case I &64.8$\pm 7.2$ &0.195$\pm 0.022$ & 62.5$\pm 6.2$ & 0.192$\pm 0.022$\\ 
Case II& 64.1$\pm 6.8$ & 0.196$\pm 0.021$ & 62.2$\pm 6.8$ & 0.191$\pm
0.021$\\
Case III &58.2$\pm 11.7$ & 0.193$\pm 0.018$ & 56.0$\pm 10.0$& 0.184 $\pm
0.014$\\
\hline\hline
\end{tabular}
\end{center}
\end{table}

In summary, building on the precise knowledge of the volume and surface
symmetry coefficients of nuclei and the nuclear saturation density,
we have proposed a model that gives the value of the symmetry slope
parameter $L$ of infinite matter. The value of $L$ so extracted is seen
to be nearly independent of the choice of different forms of density
dependence of symmetry energy and also the choice of heavy nuclei. The
small uncertainties in the known empirical values of the volume and
surface symmetry energy and the saturation density of nuclear matter
keeps $L$ in tight bounds.  This helps in determining the nuclear
EOS more accurately around the saturation density;  prediction of the
neutron skin thickness in narrower limits may further also help in
having a better feel on the proper input in the isovector channel in
the construction of energy density functionals.

\begin{acknowledgments}
 J.N.D  acknowledges support of DST, Government of India. The 
authors gratefully acknowledge the assistance of Tanuja Agrawal
in the preparation of the manuscript.
\end{acknowledgments}

\end{document}